\documentclass[8.5pt,twoside,twocolumn]{article}
\usepackage[super,sort&compress,comma]{natbib} 
\usepackage{mhchem}
\usepackage{times,mathptmx}
\usepackage{times}
\usepackage{sectsty}
\usepackage{balance} 

\usepackage{graphicx} %eps figures can be used instead
\usepackage{lastpage}
\usepackage{amsfonts,amssymb,amsmath,bm,multirow}
\usepackage[format=plain,justification=raggedright,singlelinecheck=false,font=small,labelfont=bf,labelsep=space]{caption} 
\usepackage{fancyhdr}
\usepackage{color}

\begin{document}

\thispagestyle{plain}
\fancypagestyle{plain}{
%\fancyhead[L]{\includegraphics[height=8pt]{headers/LH}}
%\fancyhead[C]{\hspace{-1cm}\includegraphics[height=20pt]{headers/CH}}
%\fancyhead[R]{\includegraphics[height=10pt]{headers/RH}\vspace{-0.2cm}}
\renewcommand{\headrulewidth}{1pt}}
\renewcommand{\thefootnote}{\fnsymbol{footnote}}
\renewcommand\footnoterule{\vspace*{1pt}% 
\hrule width 3.4in height 0.4pt \vspace*{5pt}} 
\setcounter{secnumdepth}{5}

\makeatletter 
\def\subsubsection{\@startsection{subsubsection}{3}{10pt}{-1.25ex plus -1ex minus -.1ex}{0ex plus 0ex}{\normalsize\bf}} 
\def\paragraph{\@startsection{paragraph}{4}{10pt}{-1.25ex plus -1ex minus -.1ex}{0ex plus 0ex}{\normalsize\textit}} 
\renewcommand\@biblabel[1]{#1}            
\renewcommand\@makefntext[1]% 
{\noindent\makebox[0pt][r]{\@thefnmark\,}#1}
\makeatother 
\renewcommand{\figurename}{\small{Fig.}~}
\sectionfont{\large}
\subsectionfont{\normalsize} 

\fancyfoot{}
%\fancyfoot[LO,RE]{\vspace{-7pt}\includegraphics[height=9pt]{headers/LF}}
%\fancyfoot[CO]{\vspace{-7.2pt}\hspace{12.2cm}\includegraphics{headers/RF}}
%\fancyfoot[CE]{\vspace{-7.5pt}\hspace{-13.5cm}\includegraphics{headers/RF}}
\fancyfoot[RO]{\footnotesize{\sffamily{1--\pageref{LastPage} ~\textbar  \hspace{2pt}\thepage}}}
\fancyfoot[LE]{\footnotesize{\sffamily{\thepage~\textbar\hspace{3.45cm} 1--\pageref{LastPage}}}}
\fancyhead{}
\renewcommand{\headrulewidth}{1pt} 
\renewcommand{\footrulewidth}{1pt}
\setlength{\arrayrulewidth}{1pt}
\setlength{\columnsep}{6.5mm}
\setlength\bibsep{1pt}

\twocolumn[
  \begin{@twocolumnfalse}
\noindent\LARGE{Percolation and orientational ordering in systems of magnetic nanorods}
\vspace{0.6cm}

\noindent\large{\textbf{Carlos E. Alvarez$^{\ast}$\textit{$^{a}$} and
Sabine H. L. Klapp\textit{$^{a}$}}}\vspace{0.5cm}
%Please note that \ast indicates the corresponding author(s) but no footnote text is required. 

\noindent\textit{\small{\textbf{Received Xth XXXXXXXXXX 20XX, Accepted Xth XXXXXXXXX 20XX\newline
First published on the web Xth XXXXXXXXXX 200X}}}

\noindent \textbf{\small{DOI: 10.1039/b000000x}}
\vspace{0.6cm}
%Please do not change this text.

\noindent \normalsize{
Based on Monte Carlo (MC) computer simulations we study the structure formation of a system of magnetic nanorods. Our model particles consist of fused spheres with 
permanent magnetic dipole moments, as inspired by recent experiments. The resulting system behaves significantly different from a system of hard (non-magnetic) rods or magnetic rods 
with a single longitudinal dipole. In particular, we observe for the magnetic nanorods a significant decrease of the percolation threshold (as compared to non-magnetic rods) at low 
densities, and a stabilization of the high-density nematic phase. Moreover, the percolation threshold is tunable by an external magnetic field.
%The abstract should be a single paragraph which summarises the content of the article. Any references in the abstract should be written out in full \textit{e.g.} [Surname \textit{et al., Journal Title}, 2000, \textbf{35}, 3523].
}
\vspace{0.5cm}
 \end{@twocolumnfalse}
  ]

\section{Introduction}
%Footnotes
%\footnotetext{\dag~Electronic Supplementary Information (ESI) available: [details of any supplementary information available should be included here]. See DOI: 10.1039/b000000x/}

%Please use \dag to cite the ESI in the main text of the article.
%If you article does not have ESI please remove the the \dag symbol from the title and the above footnotetext.

\footnotetext{\textit{$^{a}$~Institut f\"ur Theoretische Physik, TU Berlin, Hardenbergstra{\ss}e 36, D-10623 Berlin, Germany.}}
\footnotetext{* E-mail: carlos.e.alvarez@tu-berlin.de}

%additional addresses can be cited as above using the lower-case letters, c, d, e... If all authors are from the same address, no letter is required

%\footnotetext{\ddag~Additional footnotes to the title and authors can be included \emph{e.g.}\ `Present address:' or `These authors contributed equally to this work' as above using the symbols: \ddag, \textsection, and \P. Please place the appropriate symbol next to the author's name and include a \texttt{\textbackslash footnotetext} entry in the the correct place in the list.}

Magnetic nanoparticles \cite{Dormann} play nowadays a key role in a number of technological contexts from storage media and design of new functional materials \cite{Biermann,Evans,Jung,Shevchenko,Kaiser}, 
to medical applications \cite{Jurgons}. In many cases, the magnetic particles are suspended in a non-magnetic carrier liquid such as water or oil, yielding
a colloidal suspension often called "ferrofluid". The most prominent
ferrofluids involve {\em spherical} particles with typical sizes of about 10 nanometers. These particles
can be considered as single-domain ferromagnets; thus they have permanent magnetic dipole moments. 
The resulting anisotropic and long-range dipole-dipole interactions between the spheres 
play an important role for their cooperative behavior. Indeed, already in zero field (and small packing fractions) the energetic preference of head-to-tail configurations can lead to the formation of long chains and percolating (i.e., system-spanning) networks \cite{WeisLevesque1,Klokkenburg}, 
while at larger packing fractions, various ordered structures
are observed \cite{DeBell}. Application of an external (static) magnetic field 
on suspensions of magnetic spheres yields the formation of aligned chains and bundles \cite{Jordanovic,Heinrich}. 
As a "byproduct" 
one observes drastic changes of the material properties, particularly the shear viscosity ("magnetoviscous effect") and the thermal conductivity
\cite{Philip}. Thus, magnetic suspensions are prime examples of complex fluids, whose internal structure, 
phase behavior and dynamic rheological properties can be efficiently controlled by external parameters \cite{Ferro,Odenbach}.

Within this area of research, much effort is currently devoted to the synthesis and characterization 
of magnetic particles with {\em anisotropic} shape, examples being magnetic rods \cite{Birringer,Jemaire,Vroege,Qi} and cubes \cite{Glotzer}. One key issue, e.g., in the case of rods, is to stabilize the chains (against shear) and thus, to enhance the magnetoviscous effect observed in "simple" ferrofluids. Moreover, anisotropic magnetic particles enable per definition
a broader variety of self-assembled structures and patterns. However, compared to the case of magnetic dipolar spheres, the collective behavior
of anisotropic magnetic particles is much less understood.

In the present article, we investigate structure formation phenomena in suspensions of a special class of magnetic nanorods (MNR) by Monte Carlo (MC) computer simulations.  Our model MNRs consist of dipolar spheres
which are permanently linked ("fused") into a stiff chain with internal head-to-tail alignement of the dipole moments.
This model is inspired by recent experiments
of Birringer {\em et al.} \cite{Birringer} who used a self-assembly (aerosol-synthesis) process to produce magnetic rods composed of aligned ironoxide spheres. 
The magnetic field created by one of these rods is a superposition of the dipolar fields of the individual spherical particles. This is in contrast
to earlier models of rod-like particles with {\em single} longitudinal (or transversal) dipole moment. Indeed, models of that type have already been intensely
studied more than a decade ago, both by theory and by simulations
\cite{ZarragoicoecheaWeisLevesque,WeisLevesqueZarragoicoechea,Williamson,GilJacksonMcGrother1}.
At that time, interest was mainly driven by the desire to understand the phase behavior of polar liquid crystals. However, as we will see in the present study,
the structural behavior of  our new model MNRs differs strongly from that of single-dipole rods.

One main goal of our study is to explore the {\em percolation} behavior of the MNRs. This is an interesting issue not only in the area of
magnetic fluids (recall that already simple magnetic spheres display pronounced chain and network formation \cite{WeisLevesque1,Klokkenburg}).
Indeed, the question of percolation is intensively discussed also for general rod-like, colloidal particles, including prominent examples such as carbon nanotubes \cite{Martel,Xueshen,KyrylyukSchillingvdSchootetal}.
Research in this directions is generally driven by the desire  to produce {\em lightweight} and, at the same time, 
highly connected materials with mechanical, thermal or electrical properties that are enhanced relative to their counterparts in the corresponding non-connected systems \cite{Coleman,Deng,Salvetat,Birringer,Philip}. Thus, a general aim of these efforts is explore conditions under which
the {\em percolation threshold}, i.e., the volume fraction required for the formation of systems-spanning clusters,
 is decreased \cite{BugSafranGrestWebman}. It has already been shown that 
 such as decrease can be realized by an increase of the aspect ratio \cite{Ambrosettietal} as well as by other (interaction-related) factors such as depletion effects \cite{Vigolo,SchillingJungblutMiller,KyrylyukSchillingvdSchootetal}.  In the present study we show that
 the superpositioned magnetic interactions between our MNRs 
 provide yet another mechanism that strongly enhances 
 the tendency for percolation.

Further topics of this article concern the stability of long-range orientational order, as well as the impact of an external magnetic field \cite{Fang}. In particular, we show that 
the MNR interactions tend to {\em stabilize} the nematic phase in sufficiently long MNRs. However, contrary to what has been found in a recent analytical study of 
non-magnetic rods \cite{vdschoot}, the nematic transition does not surpress the percolation. Finally, we briefly report on the impact of an external magnetic field. It turns
out that already small magnetic fields lead to a significant decrease of the percolation threshold.

The rest of this paper is organized as follows. In section~\ref{model} we begin by defining our MNR model, followed by the details of the MC simulations in section~\ref{MCsim}. In section~\ref{results} we present our numerical results, starting with the effect of the dipolar interaction on the cluster formation.
We then proceed to a discussion of the percolation transition and the appearance of global orientational order in sections~\ref{percolation}
and \ref{density}, respectively.  In section~\ref{extField} we show our results in presence of an external magnetic field. 
Our conclusions are summarized in section~\ref{concl}.

%\subsection{This is the subsection heading style}
%Section headings can be typeset with and without numbers.
%
%\subsubsection{This is the subsubsection style.~~} These headings should end in a full point.  
%
%\paragraph{This is the next level heading.~~} For this level please use \texttt{\textbackslash paragraph}. These headings should also end in a full point.%
%

\section{Model and Simulations}\label{ModSim}

\subsection{Model}\label{model}
\begin{figure}[h]
  \centering
  \includegraphics[width=\columnwidth]{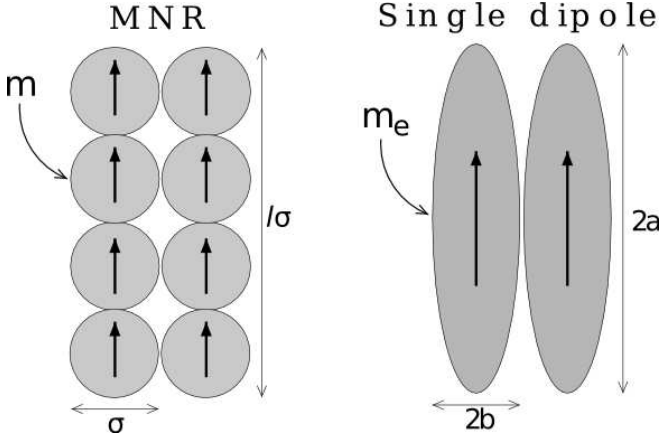} 
  \caption{Parallel side-by-side configurations of two MNRs (left) and two single-dipole spheroids (right).}
  \label{fig:models}
\end{figure}

Our model system consists of stiff nanorods made out of several identical hard spheres of diameter $\sigma$ with a 
point dipole at their center (see Fig.~\ref{fig:models} left). The positions of the spheres are fixed with 
respect to the center of mass of the rod, and the orientations of the dipoles are always aligned with the symmetry axis of the rod.
We will use $l$ to denote the number of magnetic spheres composing a nanorod,
which is also its length in units of $\sigma$, therefore $l=1$ corresponds to the dipolar hard sphere (DHS) case. 
%In this work sometimes we refer collectively to the particles as rods, including the case $l=1$.
The interaction energy between two rods $i$ and $j$ is then the sum of the pair interactions between their respective interaction 
sites (magnetic spheres). That is
\begin{equation}
u_{ij}(\bm{R}_{ij},\Omega_i,\Omega_j)=\sum_{\alpha=1}^{l_i}\sum_{\beta=1}^{l_j}
w(\bm{r}_i^{\alpha}-\bm{r}_j^{\beta},\bm{m}_i^{\alpha},\bm{m}_j^{\beta}),
\label{eq:rodInter}
\end{equation} 
where $\bm{R}_{ij}$ is the vector joining the centers of rods $i$ and $j$ with orientations $\Omega_i$ and $\Omega_j$.
Each rod is composed of $l_i$ interaction sites, with $\bm{r}_i^{\alpha}$ denoting the position of site $\alpha$ of 
particle $i$, and $\bm{m}_i^{\alpha}$ is its magnetic moment. The interaction between sites is given by
\begin{equation}
w(\bm{r},\bm{m}_i^{\alpha},\bm{m}_j^{\beta})=
\left\{\begin{array}{ll}
\infty & \text{if}\ r<\sigma \\
\left[(\bm{m}_i^{\alpha}\cdot\bm{m}_j^{\beta}-3(\bm{m}_i^{\alpha}\cdot\hat{\bm{r}})(\bm{m}_j^{\beta}\cdot\hat{\bm{r}})\right]/r^3 & 
\text{if}\ r>\sigma
\end{array}\right.,
\label{eq:dipInter}
\end{equation}
with $\bm{r}=\bm{r}_i^{\alpha}-\bm{r}_j^{\beta}$.
The magnetic moment has the same strength for each sphere and is given by
$\tilde{m}=|\bm{m}_i^{\alpha}|=\mu_r v_{sph}M_{sph}$, where $\mu_r$ is the relative magnetic permeability of the solvent, and $v_{sph}$ and $M_{sph}$
are the volume of the particles and the magnetization, respectively. These parameters can, in principle, be extracted from experiments. Here we rather
consider the reduced magnetic dipole moments $m=\sqrt{\beta\tilde{m}^2/\sigma^3}$, where $\beta=(k_{B}T)^{-1}$ with $k_B$ and $T$ being Boltzmann's constant 
and the temperature, respectively. Common experimental values for $m$ (at room temperature) are of the order $1\lesssim m\lesssim 10$ \cite{Birringer,CuevasFaraudoCamacho}.
We also consider the effect of an constant external field 
($\tilde{\bm{B}}$) parallel to the $\hat{z}$ axis of the simulation box. The interaction energy between the external field and a 
MNR is 
\begin{equation}
u_{ext, i}=-\sum_{\alpha=1}^{l_i}\bm{m}_i^{\alpha}\cdot\tilde{\bm{B}}.
\label{magpot}
\end{equation}
The coupling strength (relative to $k_BT$) then follows as $mB=\beta\tilde{m}\tilde{B}$. This expression suggests to define
\begin{equation}
B=\sqrt{\beta\sigma^3}\tilde{B}.
\end{equation}

Experimental values for the magnetic fields are of the order of 0.1 Tesla (implying
$B\sim5$ at room temperature with $\sigma=10nm$). MNR ferrofluids have been found to be susceptible already
to small fields of $\tilde{B}\lesssim10m$T (see Ref.~\cite{Birringer}), implying $B\lesssim0.5$.

In the present study we also compare the structure of the magnetic nanorods consisting of $l$ spheres (our MNRs)
to the simpler, and well studied\cite{ZarragoicoecheaWeisLevesque,ZarragoicoecheaWeisLevesque1,Satoh,Houssa,Satoh1}, model of prolate hard 
spheroids with length $2a$, width $2b$, and a single longitudinal point dipole located at the center ("single dipole model"). We set the width 
to $2b=\sigma$. Thus, the shape and volume of a spheroid is comparable to that of a MNR if $2a=l\sigma$.
In order to compare the dipolar coupling strengths within the two models, we require that the pair interaction energies in the parallel side-by-side 
configuration depicted in Fig.~\ref{fig:models} are the same. This is
\begin{equation}
u_{ij}^{(MNR)}=u_{ij}^{(sing-dip)}
\end{equation}
Using the previous condition, and the expressions for the interaction energies
\begin{equation}
u_{ij}^{(MNR)}=m^2\left[l+\sum_{n=1}^{l-1}2(l-n)\frac{1-2n^2}{(1+n^2)^{5/2}}\right],
\end{equation}
and
\begin{equation}
u_{ij}^{(sing-dip)}=m_e^2,
\end{equation}
we find that the "equivalent" dipole moment $m_e$, as defined above, is given by
\begin{equation}
m_e=m\ \left[l+\sum_{n=1}^{l-1}2(l-n)\frac{1-2n^2}{(1+n^2)^{5/2}}\right]^{1/2}.
\label{eqdip}
\end{equation}

\subsection{Monte Carlo simulations}\label{MCsim}
We carried out MC simulations in the NVT ensemble with periodic boundary conditions for monodisperse systems 
of rods with lengths $l=4$ and $l=10$, at several packing fractions $\eta$ ($\eta=lNv_{sph}/L^3$, where $L$ is the length of the simulation box).
Dipolar long range interactions were taken into account by using Ewald summations with a conducting boundary \cite{LeeuwPerramSmith}.

Simulations were made with $N=1200$ rods and $N=480$ rods for $l=4$ and $l=10$ respectively. In order to compare 
with the DHS case, we also simulate systems with $l=1$ and $N=1000$ particles. The points in phase space studied were averaged 
for $5\times10^4$ to $2\times10^5$ MC cycles, depending on the strength of the interaction. 
This relatively small number of steps is sufficient for a first, exploratory study, which is the aim of this paper.
To speed up the sampling of the phase space cluster moves and inversion moves (changing the sign 
of the orientation of the particle) were used in addition to single particle moves. Cluster algorithms that speed up the 
sampling of the phase space in Monte Carlo simulations are widely used \cite{OrkoulasPanagiotopoulos,CaillolWeisLevesque,
PriezjebPelcovitz,Linse,LiuLuijten,WhitelamGeissler,MehrabiSahimi,Almarza}. Here we define the clusters using a simple proximity criteria 
approach, in which two particles are considered to be "bonded" (and part of the same cluster) if the nearest distance between their 
surfaces is less than some value $\delta$ (see below).

The same cluster definition (which is based on a geometric, rather than an energetic criterion) is used to study the aggregation 
properties of the system. One important quantity in this context is the degree of polymerization \cite{Das,Schmidle}, defined as
\begin{equation} 
\Phi=\left<\frac{N_{cl}}{N}\right>,
\label{degpol}
\end{equation} 
where $N_{cl}$ is the number of rods that belong to a cluster composed of more than one rod and $N$ is the total number of
rods. 
Second, we consider the percolation probability $\Pi$  defined as the probability of finding at least one "infinite" cluster, that is, a cluster
connected to its own periodic images.

For both properties, $\Phi$ and $\Pi$, the choice of the parameter $\delta$ is in general arbitrary. From a physical point of view
$\delta$ may be thought of as the "hopping" distance for an excitation to go from the surface of one particle to another \cite{BugSafranGrestWebman}. Depending 
on the microscopic nature of this excitation, the hopping can be relevant for, e.g., the thermal or electrical conductivity of the material. 
Here we consider $\delta$ essentially as an adjustable parameter. Specifically, following previous studies on percolation of nanorods 
\cite{Vigolo,SchillingJungblutMiller,KyrylyukSchillingvdSchootetal}  we use a value for $\delta$ which is small with respect to the dimensions 
of the particles ($\delta=0.1\sigma$). In addition, to get a somewhat less arbitrary measure of percolation, we also calculate the geometrical 
"critical" distance $\delta_c$. The latter is defined as the averaged {\em minimum} value of $\delta$ for which an infinite cluster appears 
\cite{BugSafranGrestWebman,Ambrosettietal}. From a physical point of view, $\delta_c$ can be interpreted as an inverse measure for the conductivity 
of the system \cite{Ambrosettietal}.

For studying the structure of the system we computed some of the coefficients $h^{mnl}(r)$ of the expansion of the pair correlation 
function in terms of rotational invariants \cite{Blum}. Here we present results for the projection \cite{WeisLevesque2}
\begin{equation} %No 4Pi factors because there are no \delta(\bm{m}_i-\bm{m}) functions!
h^{220}(r)=\frac{5}{4\pi\rho^* r^2 N}\left<\sum_{i=1}^{N-1}\sum_{j>i}^N\delta(r-R_{ij})\ P_2(\cos\theta_{ij})\right>,
\label{correl}
\end{equation}
where $\rho^*=lv_{sph}\sigma^3/\eta$ is the reduced density, $P_2$ is the  Legendre polynomial of degree $2$,
and $R_{ij}$ and $\theta_{ij}$ are the center-to-center distance and the angle between the orientations of rods $i$ and $j$.

Finally, to investigate the degree of global orientational order we compute the degree of parallel ordering ("polarization") and the nematic order parameters.
The "polarization" is defined as
\begin{equation}
G_1=\left<\frac{1}{N}\left|\sum_i^N\hat{\bm{m}}_i\cdot\hat{\bm{d}}\right|\right>,
\label{ferrom}
\end{equation} 
with $\hat{\bm{d}}$ denoting the unit eigenvector associated with the largest eigenvalue of the matrix
\begin{equation}
Q_{kl}=\frac{1}{2N}\sum_{i=1}^N(3\hat{m}_k^i\hat{m}_l^i-\delta_{kl}),
\end{equation}
where the $i$ denotes the particle and the indexes $k$ and $l$ denote the cartesian component of the orientation vector.
The nematic order parameter ($G_2$) is defined as the largest eigenvalue of $Q$.
When applying an external field to the systems we measure additionally the magnetization of the system, defined as 
\begin{equation}
\bm{\mbox{M}}=\frac{\left<\sum_i^N\sum_{\alpha}^{l_i}\bm{m}_i^{\alpha}\right>}{L^3}.
\label{magne}
\end{equation}

\section{Results}\label{results}

\subsection{Cluster formation}\label{structure}
Our first goal is to investigate the type of clusters formed in our MNR system at low packing fractions $\eta$, as well as the
dependence of the clustering on the interaction strength ($m$). 
The appearance of clusters in systems of MNRs is already suggested by the
well-studied system of DHS (corresponding to the case $l=1$ in our model). Indeed, DHS particles are known
to form long, head-to-tail chains at packing fractions $\eta\lesssim 1$ 
and dipole moments $m>2$ \cite{WeisLevesque}.

Taking the DHS system as a reference, we here present results from MC simulations of rods with lengths $l=1$, $4$, and $10$ and dipole moments
$m=0$, $1.5$, and $2.4$, at a packing fraction $\eta=0.0524$. This packing fraction is comparable with typical experimental values
($\eta\approx 0.01-0.02$) for MNR systems \cite{Birringer}.

\begin{figure}[h]
  \centering
  \begin{tabular}{cccc}
  \hline
  \multicolumn{2}{|c}{\includegraphics[width=.45\columnwidth]{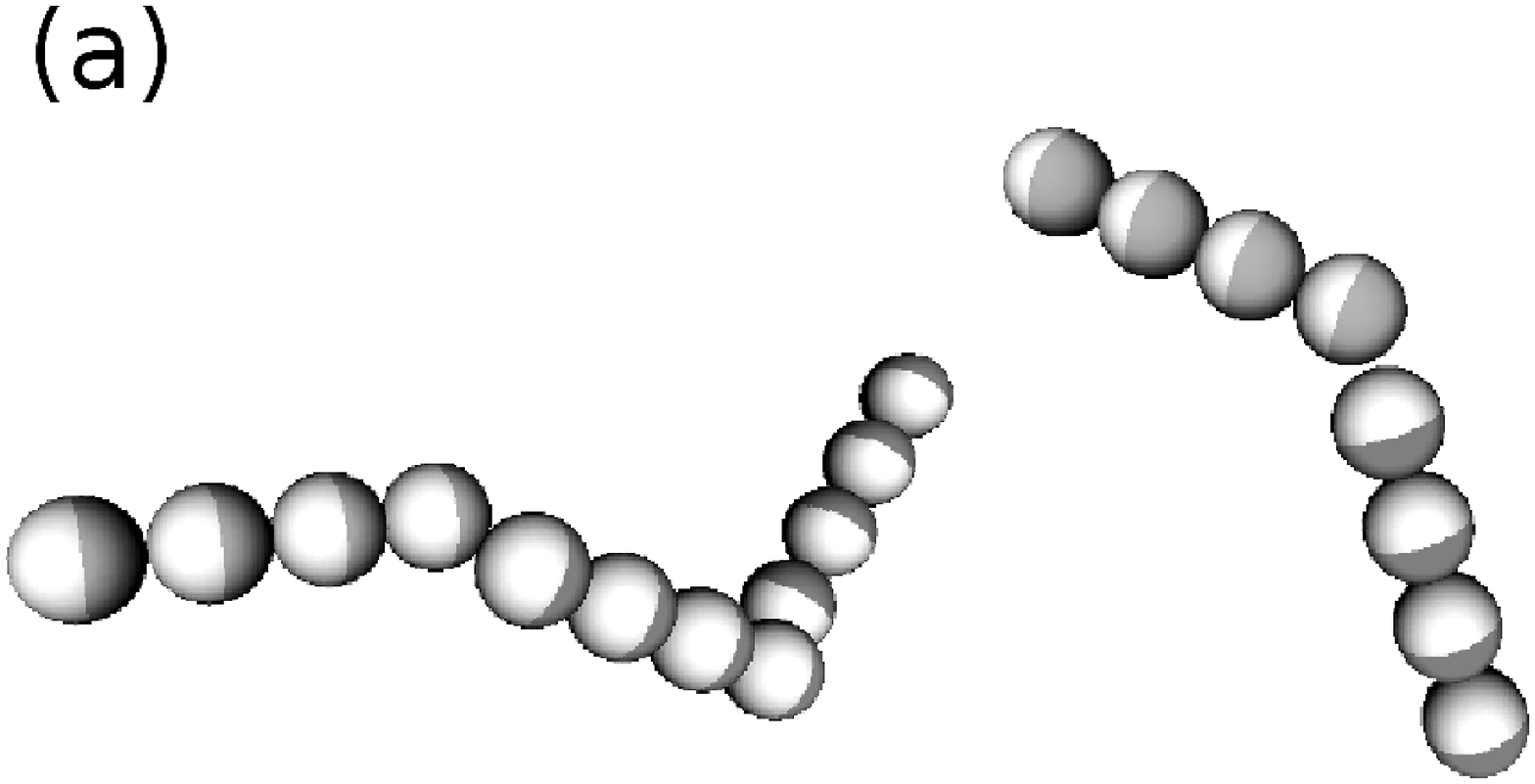}} &
  \multicolumn{2}{|c|}{\includegraphics[width=.45\columnwidth]{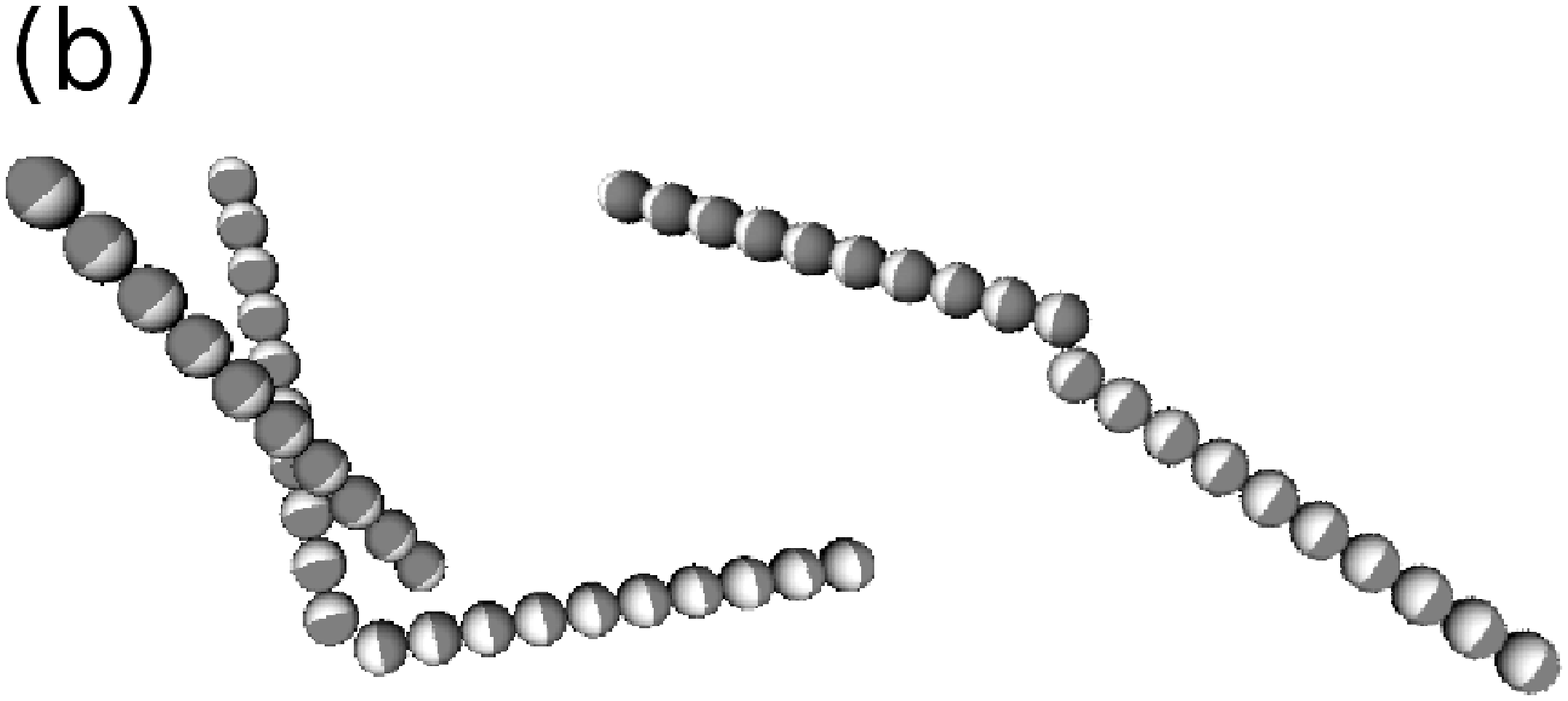}} \\
  \hline
  \multicolumn{2}{|c}{\includegraphics[width=.45\columnwidth]{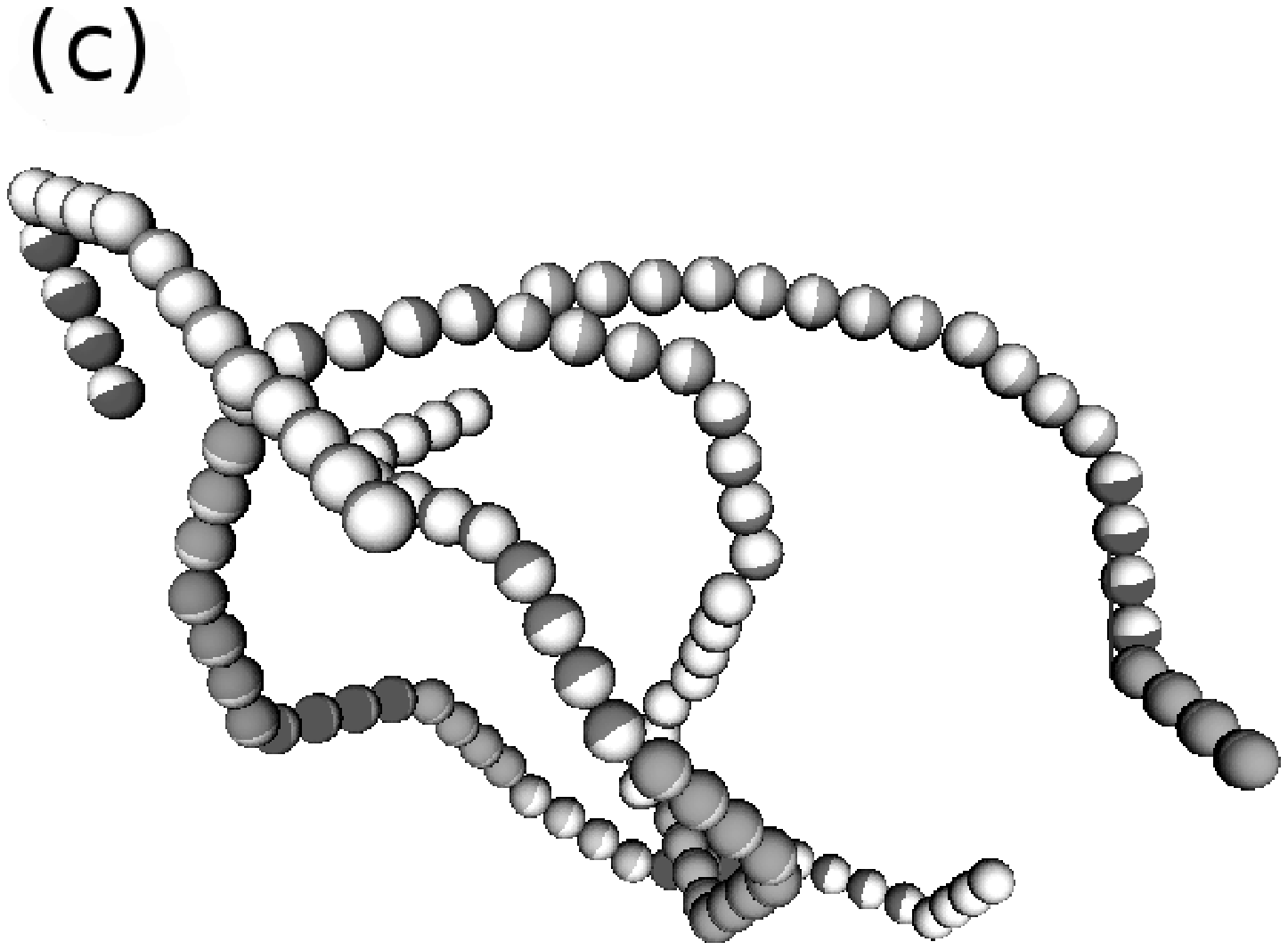}} & 
  \multicolumn{2}{|c|}{\includegraphics[width=.45\columnwidth]{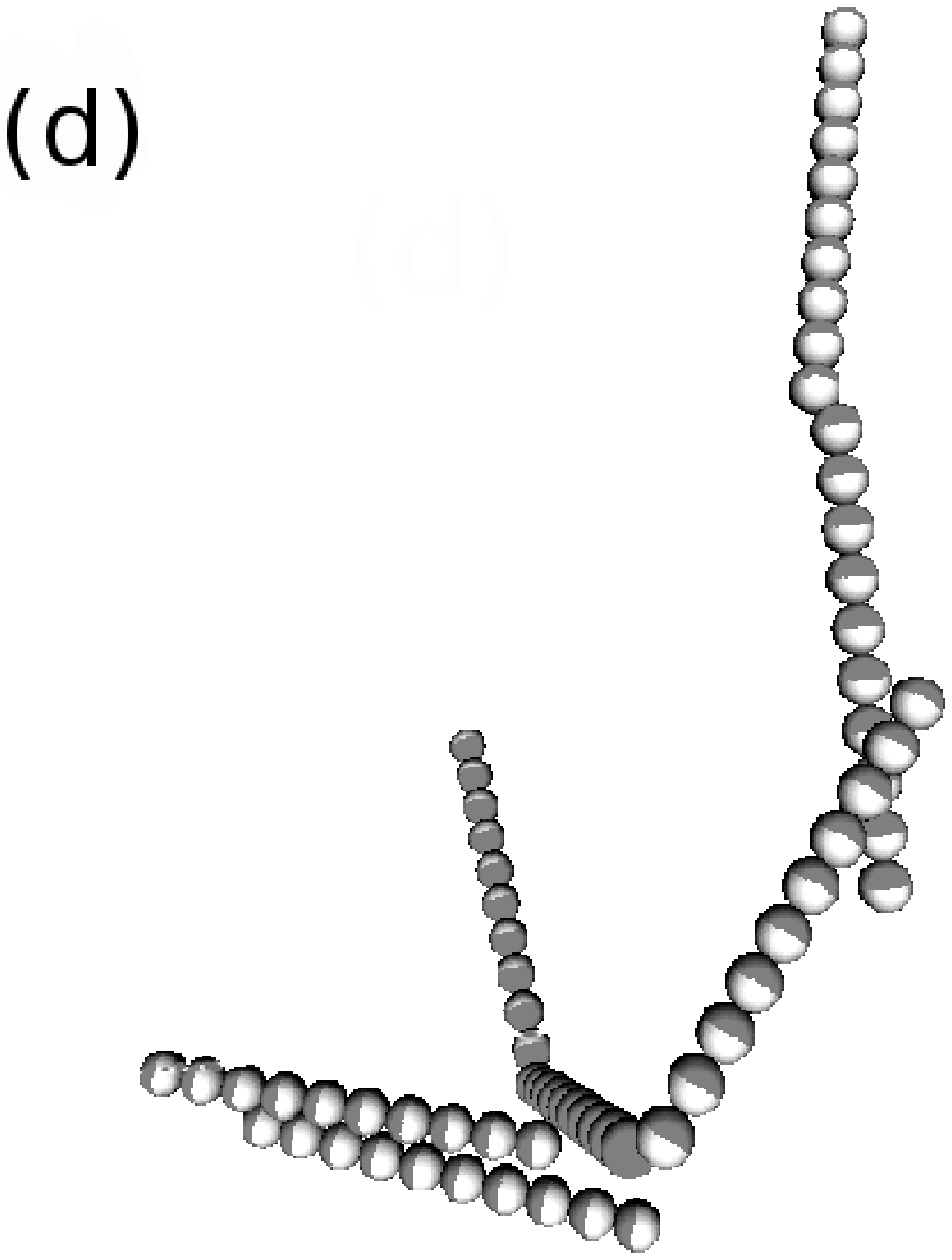}}\\ 
  \hline
  \end{tabular}
%\vspace{1cm}
  \caption{Examples of non-percolating clusters at a volume fraction $\eta=0.0524$. 
	   (a) $m=1.5$ and $l=4$. (b) $m=1.5$ and $l=10$. (c) $m=2.4$ and $l=4$. (d) $m=2.4$ and $l=10$.}
  \label{fig:cl_u_l}
\end{figure}

Corresponding MC results for the degree of polymerization [see eq.~\ref{degpol}] are given in Table~\ref{tbl:poldeg}. 
At $m=0$ (non-magnetic rods), the value of $\Phi$ is generally small and there is no formation of chains or other larger structures, regardless of the value of $l$. 
This is expected since the packing fraction considered here is far below the percolation threshold of that system (see Section 3.2). At $m=1.5$, about 30-40 percent
of the rods are associated into clusters [see Table~\ref{tbl:poldeg}] which contain, however, typically less than three particles. Some representative snapshots of aggregates
at $m=1.5$ and $l=4$, $10$ are shown in Figs.~\ref{fig:cl_u_l} (a) and (b). It is seen that the rods already tend to form head-to-tail arrangements, despite of the still 
rather small interaction strength. Finally, at $m=2.4$ essentially all rods are associated into clusters, as seen from the fact that $\Phi$ is close to one for all values of $l$ 
[see Table~\ref{tbl:poldeg}]. As expected, these clusters contain on the average more rods than those at $m=1.5$. The resulting structures are illustrated by the snapshots
in Figs.~\ref{fig:cl_u_l} (c) and (d). These snapshots also reveal that there are essentially {\em three} types of configurations 
which seem to be preferred by strongly coupled MNRs. Type~I, which is the dominant one at $l=4$ [see Fig. 2 (c)], is given by head-to-tail chains or rings, similar to what 
is found in conventional DHS systems ($l=1$). Type~II, which is seen in the $l=10$ system, and, to less extent, also in the $l=4$ system, is a side-by-side configuration where the dipoles point in opposite (antiparallel) direction. 
Finally, Type~III, which seems to be particularly important for long MNRs ($l=10$), [see Fig.~\ref{fig:cl_u_l} (d)] consists of a configuration where the dipoles are oriented along the 
same direction (parallel), and the rods are close to side-by-side, but somewhat shifted against one another in longitudinal direction.
\begin{table}[h]
\small
 \caption{\ Degree of polymerization at $\eta=0.0524$.}
 \label{tbl:poldeg}
 \begin{tabular*}{0.5\textwidth}{@{\extracolsep{\fill}}cll}
   \hline
   $l$ & $m$ & $\Phi$\\
   \hline
   1 & 0.0 & 0.147 \\
   4 & 0.0 & 0.193 \\
   10 & 0.0 & 0.303 \\
   1 & 1.5 & 0.312 \\
   4 & 1.5 & 0.349 \\
   10 & 1.5 & 0.396 \\
   1 & 2.4 & 0.964 \\
   4 & 2.4 & 0.997 \\
   10 & 2.4 & 0.954 \\
   \hline
 \end{tabular*}
\end{table}

\begin{figure}[h]
\centering
  \includegraphics[width=\columnwidth]{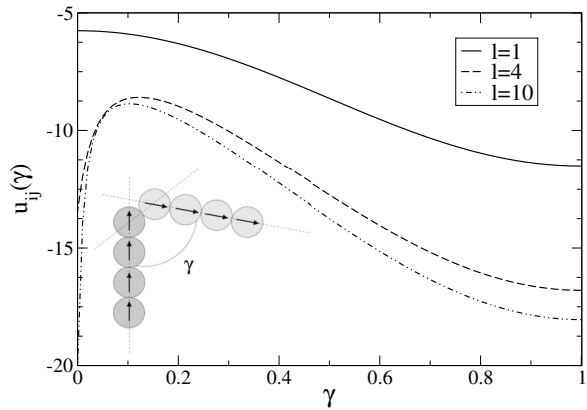}
\vspace{-1cm}
  \caption{Interaction energy between two rods ($i$ and $j$) composed of $l$ magnetic spheres 
	   with $m=2.4$, in Type I configuration as a function of the "joint" angle $\gamma$, in units of $\pi$. $\gamma=1$ is
	   the straight head-to-tail configuration and $\gamma=0$ is the antiparallel configuration.}
  \label{fig:EnervsT}
\end{figure}

To better understand these configurations we take a closer look at the energetic landscape of interacting MNRs. A first useful insight is given by Fig.~\ref{fig:EnervsT} 
which shows the interaction energy of a pair of MNRs as a function of the "joint" angle $\gamma$ (in units of $\pi$). The latter is defined such that $\gamma=1$ 
and $\gamma=0$ correspond to the Type~I and Type~II configurations, respectively. We see from Fig.~\ref{fig:EnervsT} that for DHS ($l=1$) there is just one energy minimum with 
respect to $\gamma$, which corresponds to Type~I. For $l=4$ and $l=10$, on the other hand, there is a second pronounced minimum at $\gamma=0$. Moreover, this 
second minimum is separated from the first one by a large energy barrier which becomes the larger as $l$ is increased. This suggests that both, Type~I and Type~II 
configurations are quite stable against (thermal) fluctuations, a picture which is indeed confirmed in the actual MC simulations.

\begin{figure}[h]
\centering
  \includegraphics[width=\columnwidth]{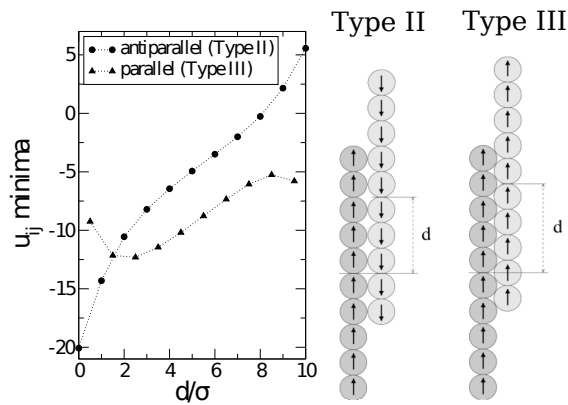}
\vspace{-1cm}
  \caption{Left: Interaction energy {\it minima} between two adjacent parallel rods of length $l=10$ with  antiparallel (Type II) and parallel (Type III) 
	   dipole moments ($m=2.4$). Right: Illustration of the configurations.}
  \label{fig:Enervsd}
\end{figure}

Further, especially for longer MNRs ($l=10$) it is interesting to compare the energies related to Type~II and Type~III configurations. This is done in Fig.\ref{fig:Enervsd}, 
where we plot the interaction energies of two adjacent rods (see sketch in the upper part of the figure) with parallel rod axes and
either antiparallel (Type~II) or parallel (Type III) orientation of the dipole moments as function of the "longitudinal" distance $d$ (i.e., the displacement
of the rod centers along the direction of the rod axes). Note that we have plotted only the {\em minima} of the interaction energies
with respect to $d$. For the antiparallel configurations, these minima occur at $d^{min}=n\,\sigma$ with $n$ being an integer, whereas for the parallel
configurations, $d^{min}=(n+1/2)\,\sigma$ corresponding to shifted (Type~III) arrangements. From the numerical values of the two interaction energies we can see
that both type of configurations are stable (in the sense that the energies are negative) for a broad range of displacements. Moreover, for larger values of $d$ the 
energies related to Type~III becomes even more attractive than those related to Type~II. Finally, for the specific values of $l=10$ and the specific lateral distances 
considered in Fig.~\ref{fig:Enervsd}, the configurations with lowest energy are an antiparallel one with $d=0$ (i.e., no displacement at all) and a parallel one with
$d=2.5\sigma$.

\begin{figure}[h]
\vspace{1.cm}
  \centering
  \includegraphics[width=\columnwidth]{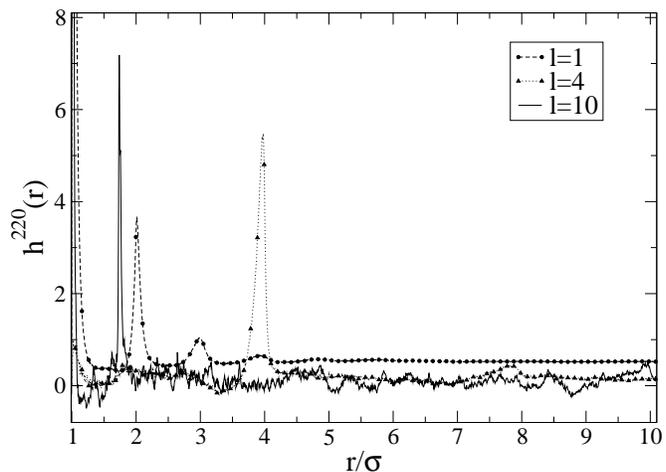} 
%\vspace{1cm}
  \caption{The correlation function $h^{220}(r)$ for MNRs with $m=2.4$ at $\eta=0.0524$.}
  \label{fig:corrhigh}
\end{figure}

The presence of several types of preferred structures in strongly coupled MNR systems is also reflected by the correlation function $h^{220}(r)$ plotted in Fig.~\ref{fig:corrhigh}. 
At $l=1$, one observes the typical peaks at multiples of a particle diameter, reflecting the (head-to-tail) alignment of neighboring particles
in the chains. The same type of (chain-like) structure is preferred by the system with $l=4$, as indicated by the large peak of $h^{220}(r)$ at $r=4\sigma$. 
Note, however, that the $l=4$ system also has a peak at $r/\sigma=1$ related to an antiparallel side-by-side (Type~II) configuration. 
At $l=10$, the highest peaks occur at $r=1\sigma$ and $r\approx \sqrt{3}\sigma$, whereas the (expected) peak at $r=10\sigma$ is already to small to be distinguished 
from statistical noise. The large heights of the first two peaks reflect the presence of both, Type~II and Type~III configurations. 

\begin{figure}[h]
  \centering
  \begin{tabular}{cccc}
  \hline
  \multicolumn{2}{|c}{\includegraphics[width=.45\columnwidth]{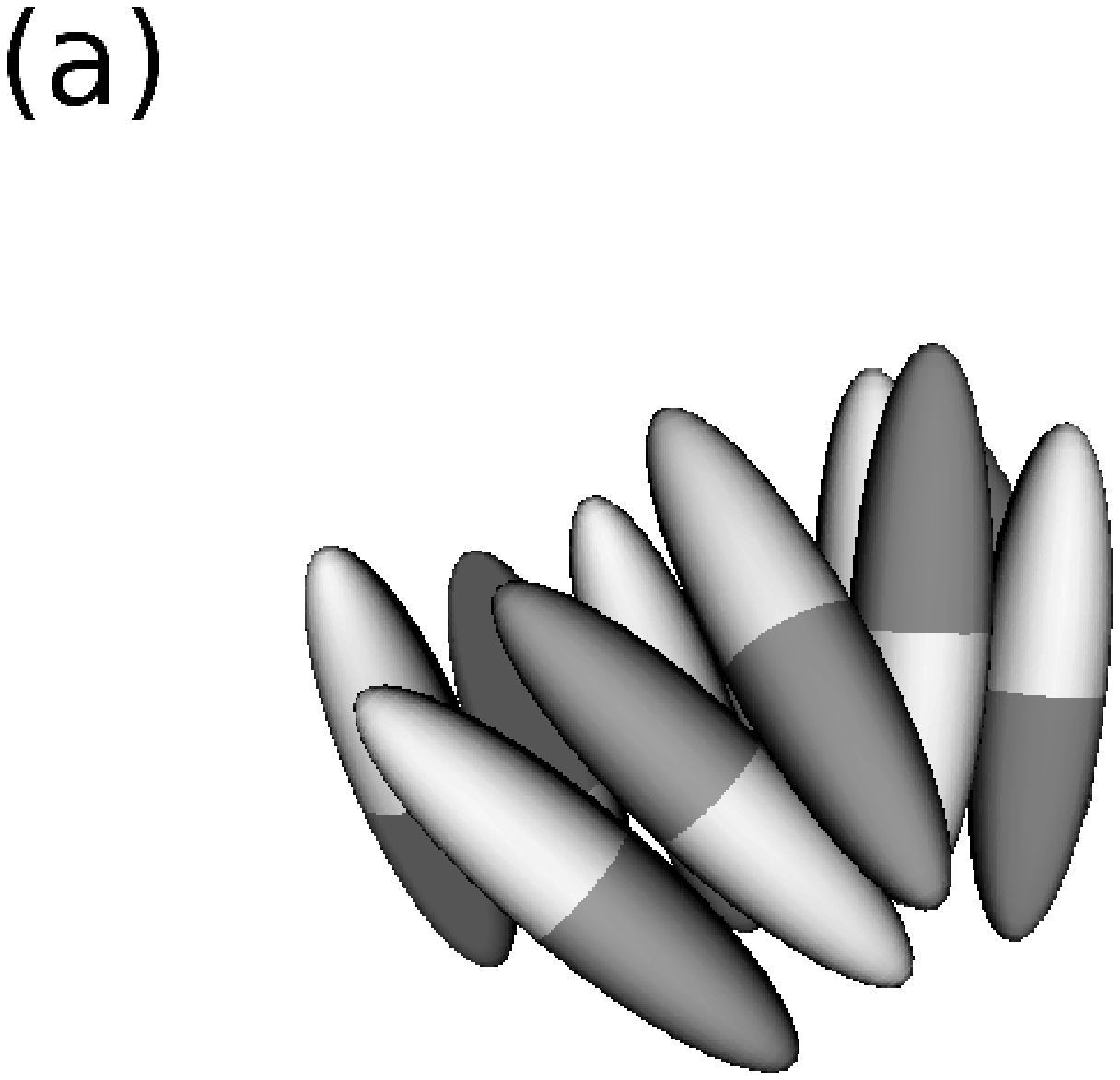}} &
  \multicolumn{2}{|c|}{\includegraphics[width=.45\columnwidth]{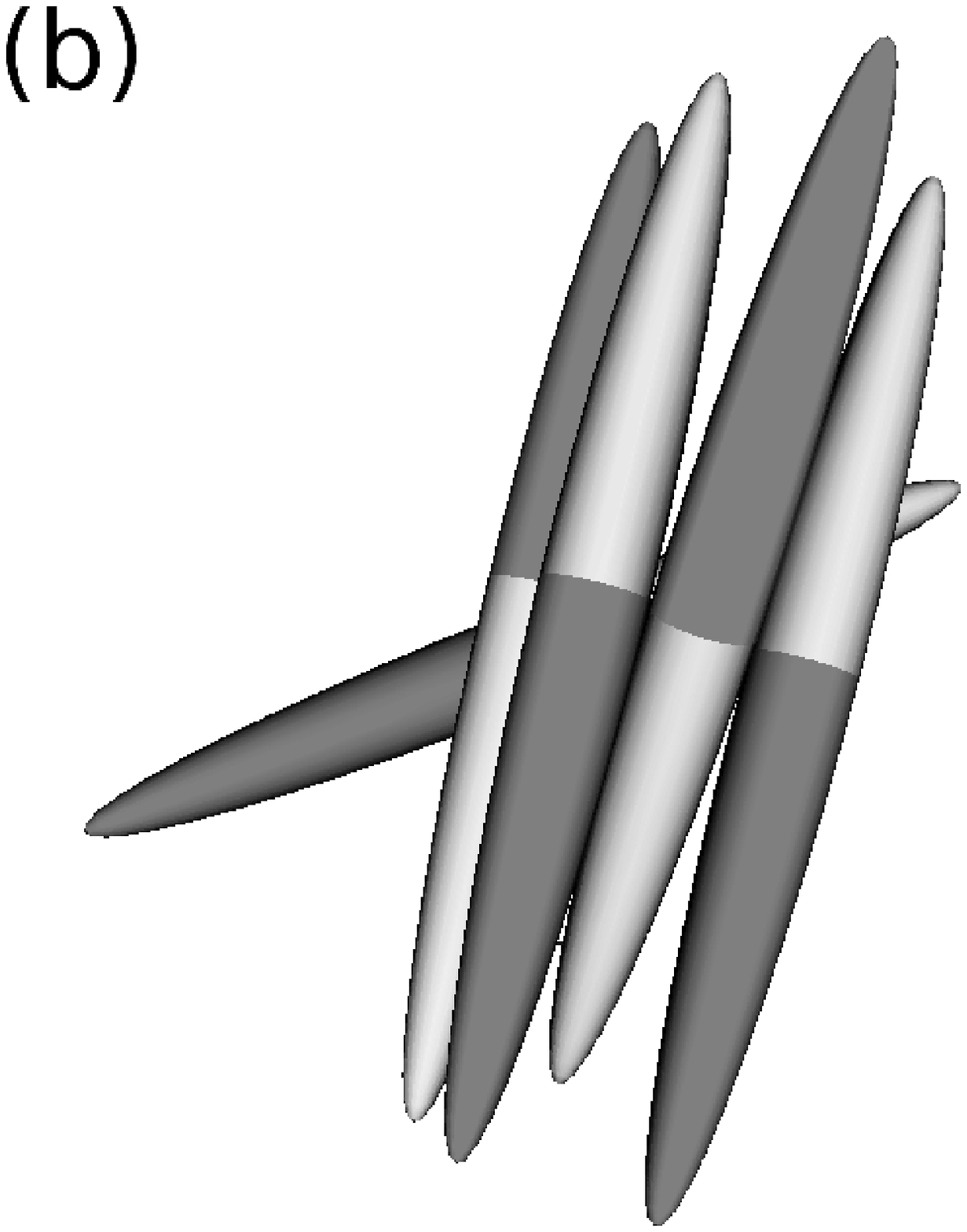}} \\
  \hline
  \end{tabular}
%\vspace{1cm}
  \caption{Sample clusters of hard spheroidal rods with a longitudinal point dipole at their center, with $m_e$ given by eq. (\ref{eqdip}) with $m=2.4$. The
	   clusters are defined with a proximity criteria using $\delta=0.2b$, and at a volume fraction $\eta=0.0524$. (a) $a/b=4$. (b) $a/b=10$.}
  \label{fig:cl_u_ar}
\end{figure}

To close this section, and to better acknowledge the specific type of clusters formed in the present model of MNRs, we show in Fig.~\ref{fig:cl_u_ar} two typical clusters formed
in systems of magnetic prolate spheroids with a single, longitudinal (point) dipole. For both aspect ratios considered in Fig.~\ref{fig:cl_u_ar}, the spheroids tend to align in 
an antiparallel side-by-side configurations, similar to the previously introduced Type~II configurations with d=0. On the other hand, configurations of Type~III are 
essentially non existent. The presence of antiparallel configurations in the single-dipole rods leads to quite compact clusters. This is consistent to findings in earlier MC studies \cite{GilJacksonMcGrother1} of such systems.

\subsection{Percolation}\label{percolation}
\begin{figure}[h]
\vspace{1.cm}
\centering
  \includegraphics[width=\columnwidth]{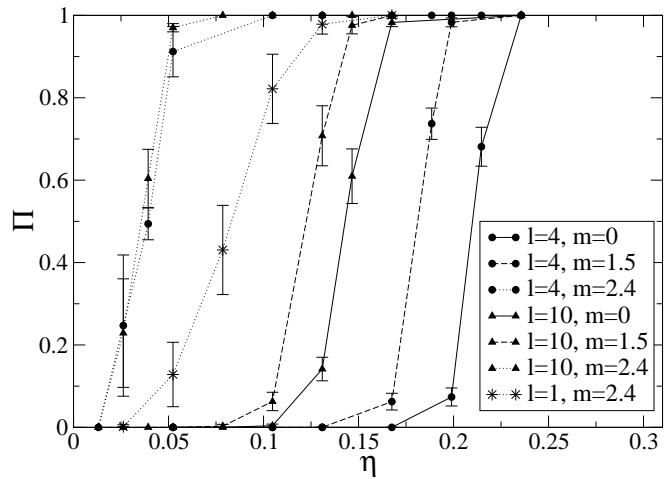}
%\vspace{1cm}
  \caption{Percolation probability as a function of the volume fraction using $\delta=0.1\sigma$.}
  \label{fig:percol}
\end{figure}

As a next step we want to understand the influence of the magnetic interaction on the percolation properties of the MNRs. It is well known
that percolation (i.e., formation of system-spanning clusters) already occurs in non-magnetic systems of prolate, hard-core particles, with the percolation 
density decreasing upon increase of the aspect ratio of the particles \cite{Leung,Ambrosettietal}. In Fig.~\ref{fig:percol} we plot our present MC results 
for the percolation probability $\Pi$ of MNRs with $l>1$ and various values of $m$ as function of the packing fraction. All data in Fig.~\ref{fig:percol}
have been obtained with a fixed value of the parameter $\delta$ determining our cluster criterion (see section \ref{MCsim}).

Considering first the non-magnetic case, we see that both the systems with $l=4$ and $10$ do have a percolation transition, as indicated
by the steep increase of $\Pi$ from values close to zero to values close to one (upon increase of $\eta$). A precise determination of the percolation transition
would require calculations for different system sizes, which we did not perform in our present, more exploratory study. Nevertheless, we can estimate
from the data in Fig.~\ref{fig:percol} that the percolation thresholds at $m=0$, using $\delta=0.1\sigma$, are at $\eta_c^{l=4}\approx0.21$ and $\eta_c^{l=10}\approx0.14$. 
To check the effect of $\delta$ we performed calculations using $\delta=0.2\sigma$ and obtained $\eta_c^{l=4}\approx0.15$ and $\eta_c^{l=10}\approx0.09$. 
When comparing the multiple-sphere rods with
spherocylinders is important to remember two things: i) The length $l$ of our rods is related to the aspect ratio of the spherocylinderts via $l=L/D+1$, 
where $L$ is the length of the cylinder and $D$ the diameter of the spherical cap of the spherocylinder.
ii) At the same aspect ratio, the volume fraction of the multiple-sphere rod is lower than that of the spherocylinder. This means we have to multiply the volume
fraction obtained for the multiple-sphere rod  by a factor $v_{spcyl}/v_{msph}$, where $v_{spcyl}$ is the volume of the spherocylinder and $v_{msph}$ is the volume of 
the multiple-sphere rod with $l=L/D+1$.
Taking into account these considerations, our results for $m=0$ are in agreement with those reported for hard spherocylinders \cite{SchillingJungblutMiller}.
Coming back to the behaviour of MNRs, we note that the trend of longer rods percolating at lower densities, persists when we "switch on" the magnetic interactions, as revealed by the data for $m=1.5$. 
More importantly, we also see that the corresponding thresholds are smaller than those for $m=0$, indicating that the magnetic interactions {\em promote} the 
percolation transition. This effect becomes even more pronounced for $m=2.4$. For the latter case, we also see that the percolation thresholds seem to saturate 
for different rod length at $\eta_c\approx 0.03$.

\begin{figure}[h]
\vspace{1cm}
  \centering
  \includegraphics[width=\columnwidth]{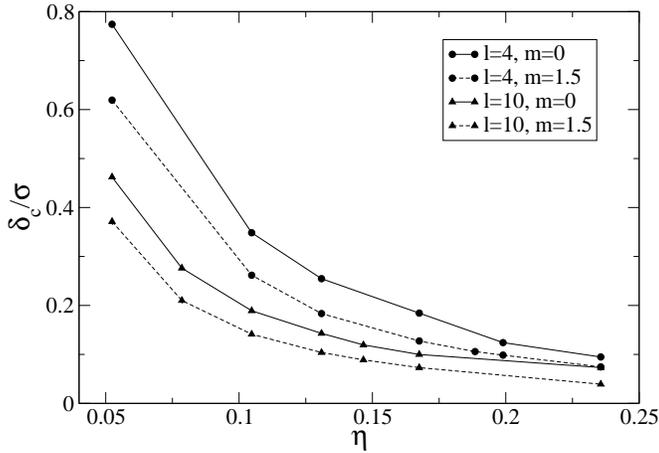}
%\vspace{1cm}
  \caption{\label{fig:dcvseta}Geometrical "critical" distance (in units of $\sigma$) vs. volume fraction for short 
  ($l=4$) and long ($l=10$) rods with ($m=1.5$) and without ($m=0$) magnetic interaction.}
\end{figure}

Having obtained the percolation transition(s) for a fixed value of the cluster parameter $\delta$, it is interesting to look at the percolation phenomenon 
from a less "biased" perspective, that is, without making any assumptions on particle distances within clusters. To this end we now consider the parameter
$\delta_c$, defined in section \ref{MCsim} as the minimum value of $\delta$ for which a percolating cluster appears in the system. 
Results for $\delta_c$ as function of $\eta$ are plotted in Fig.~\ref{fig:dcvseta}. As expected, all systems exhibit a decrease
of $\delta_c$ as the density increases. More importantly, we see that already for $m=0$, the values of $\delta_c$ for the shorter rods
($l=4$) are consistently larger than those for $l=10$, indicating that the tendency for percolation is enhanced upon increasing the rod length. 
The same trend appears when we "switch on" the magnetic interactions, as shown by the data for $m=1.5$ in Fig.~\ref{fig:dcvseta},
where the curves for fixed $l$ are shifted to the left (i.e., to lower packing fractions) relative to the corresponding ones at $m=0$. 
Thus, the magnetic interactions have the same supportive effect on the percolation.
These findings are qualitatively consistent to the ones obtained for fixed $\delta$ plotted in Fig.~\ref{fig:percol}.

Finally, it is interesting to briefly compare the percolation behavior of MNRs system to that of systems of single-dipole spheroids.
To this end, we consider the parameter
\begin{equation}
\Delta\delta_R=\frac{\delta_c-\delta_c^{ref}}{\delta_c^{ref}}\times 100,
\end{equation}
which measures, for fixed values of $\eta$ and $m$, the value of $\delta_c$ obtained for magnetic (MNR or single-dipole) particles 
relative to the corresponding ones for $m=0$ ($\delta_c^{ref}$). Results for $\Delta\delta_R$ obtained for $m=2.4$ and a low density are 
given in Table~\ref{tbl:compd}. In this context, the simulations have been started from an initial configuration with randomly positioned and
oriented rods. The systems have then been equilibrated for $\sim 3\times 10^5$ MC steps.
\begin{table}[h]
\small
  \caption{\ Comparison of $\Delta\delta_R$ for different rod models with $m=2.4$, $\eta=0.0524$ and $m_e$ given 
	  by eq. (\ref{eqdip}).}
  \label{tbl:compd}
  \begin{tabular*}{0.5\textwidth}{@{\extracolsep{\fill}}cll}
    \hline
    $l=a/b$ & $\Delta\delta_R$(MNR) & $\Delta\delta_R$(sing-dip)\\
    \hline
    4 & -87.8\% & 16.3\% \\
    10 & -83.4\% & 24.5\% \\
    \hline
  \end{tabular*}
\end{table}
We see from Table~\ref{tbl:compd} that for the MNR model the magnetic interaction produces negative values of $\Delta\delta_R$ and thus, a decrease of $\delta_c$ relative
to the non-magnetic case, consistent with the results plotted in Fig.~\ref{fig:dcvseta}. For the single-dipole model, one the other
hand, $\Delta\delta_R$ is positive, meaning that the magnetic interactions rather {\em hinder} the percolation.

This remarkable difference in the behaviors of the two models can be attributed to the different structure of the corresponding clusters. 
Whereas the MNRs tend to form open, elongated structures [see Fig.~\ref{fig:cl_u_l}], the single-dipole spheroids rather tend to compact clusters with 
antiparallel local ordering [see Fig.~\ref{fig:cl_u_ar}]. 

\subsection{Nematic ordering}\label{density}
So far we have focused on the clustering and percolation behavior of the magnetic nanorods.
However, given the non-spherical, elongated shape of the MNRs  it is tempting to consider
also the appearance of {\em long-ranged} orientational (i.e., liquid-crystalline) ordering in our model. Indeed, it is 
well established that fluids of non-spherical particles can display entropy-driven phase transitions
into nematic, smectic, and plastic phases, as the packing fraction is increased \cite{AEFM}. In particular, the phase diagram
of hard spherocylinders has been mapped out in detail by 
Bolhuis and Frenkel \cite{Bolhuis} who performed MC simulations
for a broad range of length-to-breadth ratios $L/D$ (where we recall that $L/D=l-1$). According to their results, rods with $L/D>3.7$ 
display both, a nematic and a smectic-A phase, the latter being characterized by the formation of layers in planes transversal to 
the global director. Moreover, there are MC results for {\em dipolar} hard spherocylinders with $L/D=5$ and a single longitudinal 
point dipole \cite{Williamson}. According to this study, the (longitudinal) dipole tends to {\em destabilize} the nematic phase. Instead, 
one observes an enhanced tendency for smectic ordering. The stabilization of smectic-like ordering in these systems can be understood as 
a consequence of the strong energetic preference of antiparallel side-by-side configurations. Similar observations have been made for other 
model systems involving elongated particles with longitudinal dipole moments \cite{WeisLevesqueZarragoicoechea,GilJacksonMcGrother1}.

\begin{figure}[h]
\vspace{1.cm}
\centering
  \includegraphics[width=\columnwidth]{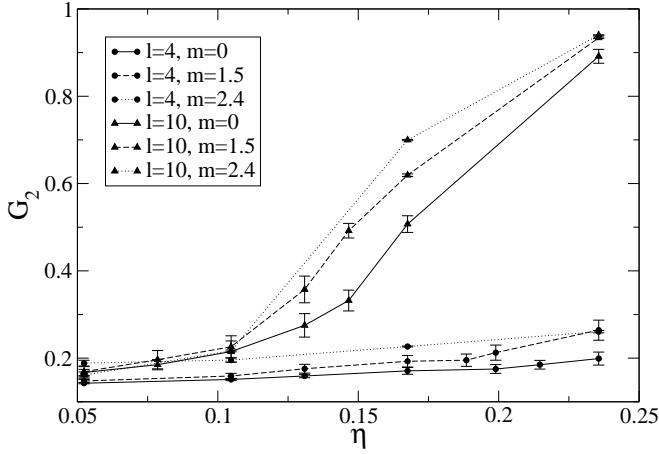}
%\vspace{1cm}
  \caption{Nematic order parameter as a function of the volume fraction for several lengths and dipole moments.}
  \label{fig:G2vsVolfrac}
\end{figure}

In the present study we investigate the occurrence of liquid-crystalline (specifically, nematic) ordering via the orientational order 
parameter $G_2$ defined in section \ref{MCsim}. Results for $G_2$ as function of the packing fraction $\eta$ are plotted in Fig.~\ref{fig:G2vsVolfrac}.

For the case $l=4$, the order parameter has negligible values regardless of the dipole moment, indicating that there is
no orientational ordering within the density range considered ($\eta\lesssim 0.25$).  Indeed, for the particular case $m=0$, this result
is consistent with the MC study of Bolhuis and Frenkel \cite{Bolhuis} who showed that hard spherocylinders with $L/D+1=4$ do not present an
isotropic to nematic phase transition.

On the other hand, we find from Fig.~\ref{fig:G2vsVolfrac} that the systems with $l=10$ clearly do exhibit ordered
phases, as reflected by the pronounced increase of $G_2$ (upon increasing $\eta$) towards values close to $1$.
At the same time, the corresponding values of the "ferromagnetic" order parameter $G_1$ defined in eq.~(\ref{ferrom}) are negligible, 
indicating that we indeed observe an ordering of the rod {\em axes} (rather than ordering of their dipole directions).
Due to the relatively small systems sizes in our simulations, the $G_2$-curves in Fig.~\ref{fig:G2vsVolfrac} are rather rounded (in fact, a thorough
determination of the packing fraction related to the ordering transition would require a systematic finite-size study). Nevertheless, we can extract two important
findings from these curves. First, considering our model rods with $l=10$ and $m=0$, the behavior of the corresponding $G_2$ appears to be consistent with 
the fact that the isotropic-nematic transition for pure hard spherocylinders (of aspect ratio $L/D+1=10$) occurs at $\eta\approx 0.25$ \cite{Bolhuis}
(indeed, taking into account the 
same considerations made in section~\ref{percolation} concerning the comparison of spherocylinders and multi-sphere rods,
the packing fraction related to the onset of nematic ordering follows from Ref.~\cite{Bolhuis} as $\eta\approx 0.17$). 
The second, and in the present context more important, finding concerns the influence of magnetic interactions. Indeed, we see that the functions $G_2(\eta)$ are shifted 
towards lower packing fractions when $m$ is increased from zero. In other words, the nematic phase is {\em stabilized} (relative to the isotropic phase) due to magnetic 
interactions. This is is strong contrast to the behavior found in systems of dipolar spherocylinders \cite{Williamson,GilJacksonMcGrother1} (and other elongated particles 
with single, longitudinal dipoles \cite{WeisLevesqueZarragoicoechea}), where the nematic state is rather suppressed.

\begin{figure}[h]
\vspace{1cm}
  \centering
  \begin{tabular}{lc}
  \includegraphics[width=.48\columnwidth]{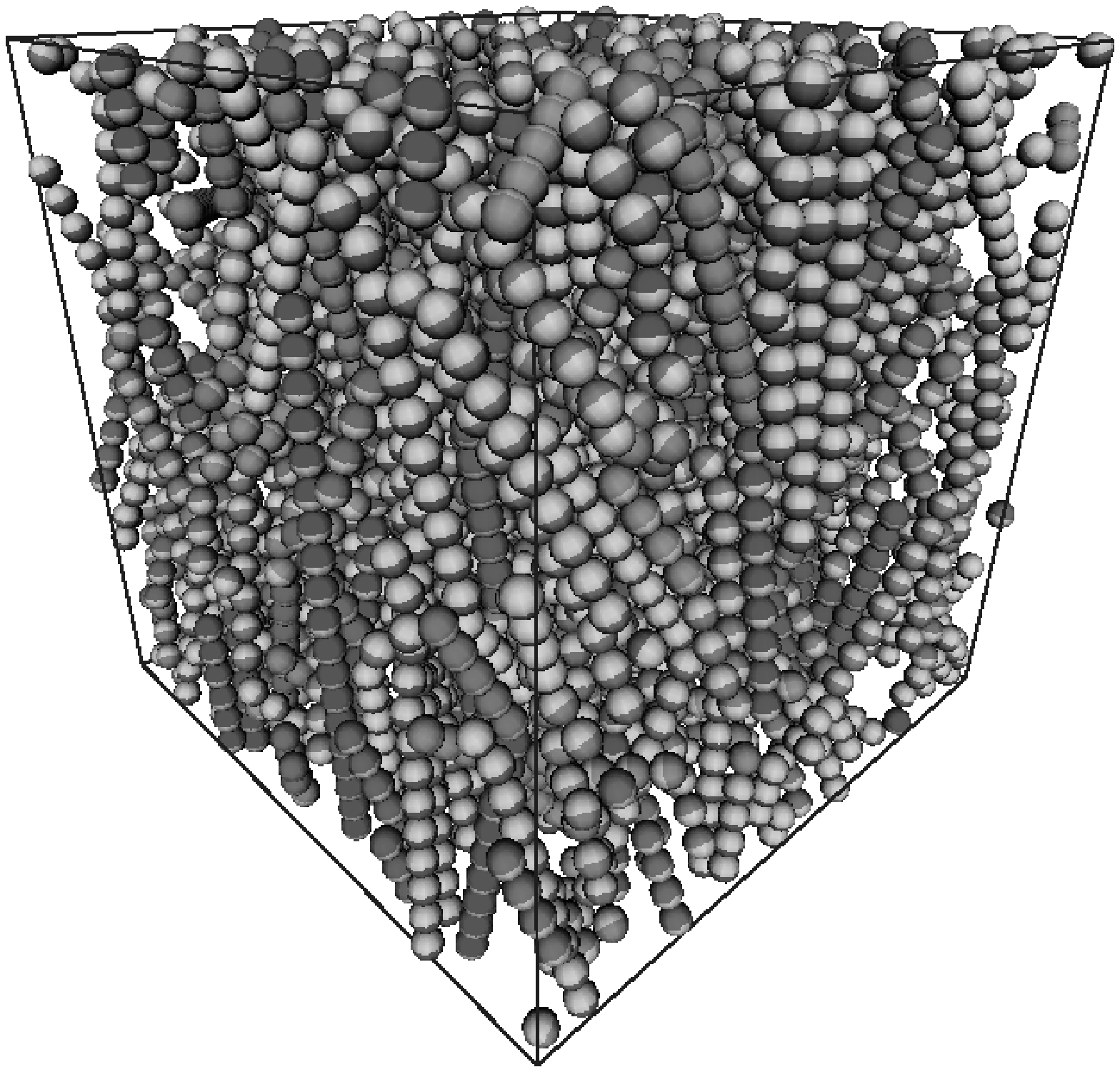} & 
  \includegraphics[height=.48\columnwidth]{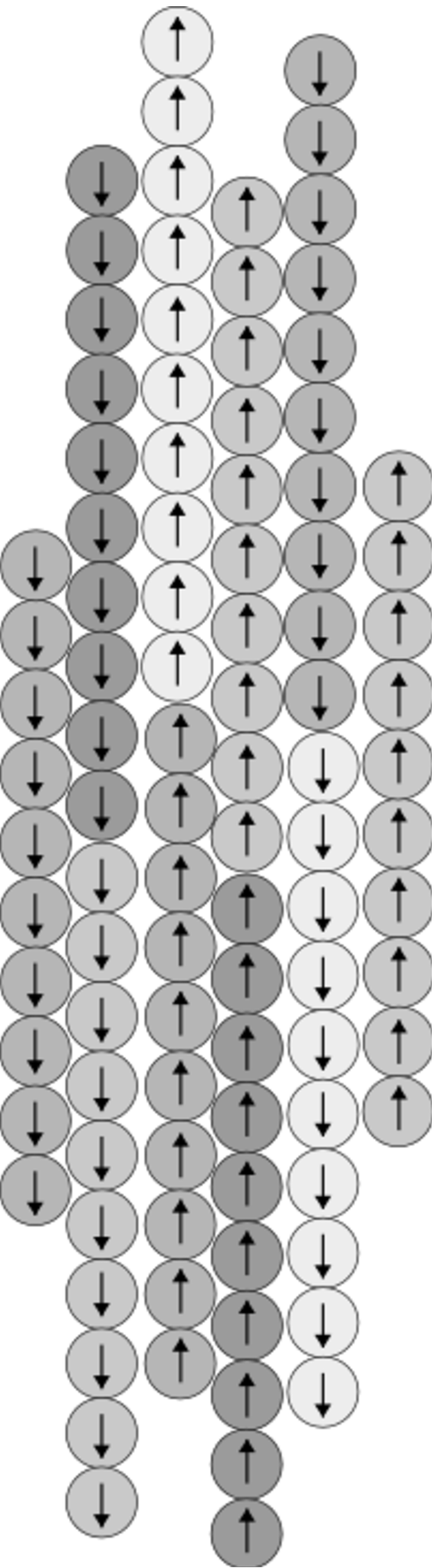}
  \end{tabular}
%\vspace{1cm}
  \caption{\ Left: Sample configuration of a system with $l=10$, $m=2.4$ and $\eta=0.1676$ ($G_2=0.699$). Right: Illustration of 
	  one way in which the magnetic interaction could stabilize the nematic phase.}
  \label{fig:nemStabil}
\end{figure}

As an attempt to understand these differences, it is useful to consider the actual structure of the present MNR systems within their nematic state.
A simulation snapshot is shown in Fig.~\ref{fig:nemStabil} (left). Closer inspection reveals that there are both, antiparallel (Type~II) configurations and shifted
parallel (Type~III) configurations, consistent with our considerations concerning the energy landscape in section \ref{structure}. 
The "coexistence" of these configurations is also illustrated by the sketch in the right part of Fig.~\ref{fig:nemStabil}.
We suspect that it is particularly the presence 
of the shifted-parallel configurations, which could stabilize the nematic against the smectic-A phase formed in dense systems of
(antiparallel oriented) single-dipolar elongated particles \cite{Williamson,GilJacksonMcGrother1}.

\subsection{Effect of an external field}\label{extField}

\begin{figure}[h]
\vspace{1cm}
  \centering
  \includegraphics[width=\columnwidth]{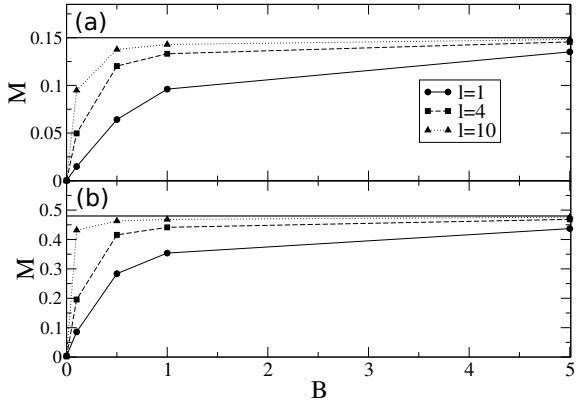}
%\vspace{1cm}
  \caption{\label{fig:MvB}Magnetization of the ferrofluid with $m=1.5$ as a function of the external field strength ($B$) 
	    for two volume fractions: (a) $\eta=0.0524$ and (b) $\eta=0.1676$.}
\end{figure}

In this last section we briefly discuss the impact of a homogeneous, external magnetic field $\bm{B}=B\,\hat{\bm{z}}$ on the 
structure of our MNR systems. We start by considering the norm of the field-induced magnetization (M=$|\bm{\mbox{M}}|$, see eq.~(\ref{magne})).
Results for the functions M$(B)$ at two packing fractions, various rod lengths and a fixed reduced dipole moment of $m=1.5$ are plotted in 
Fig.~\ref{fig:MvB}. For both values of $\eta$ considered, the longer rods are seen to be more susceptible than the shorter rods,
in the sense that the
magnetization rises more sharply with the field and reaches earlier its saturation value. 
We also see that a decrease of $B$ to zero leads to a vanishing of the magnetization in all cases considered, consistent
with our result in section~\ref{density} that the $G_1$ order parameter is zero (no spontaneous magnetization).

\begin{figure}[h]
\vspace{1.cm}
  \centering
  \includegraphics[width=\columnwidth]{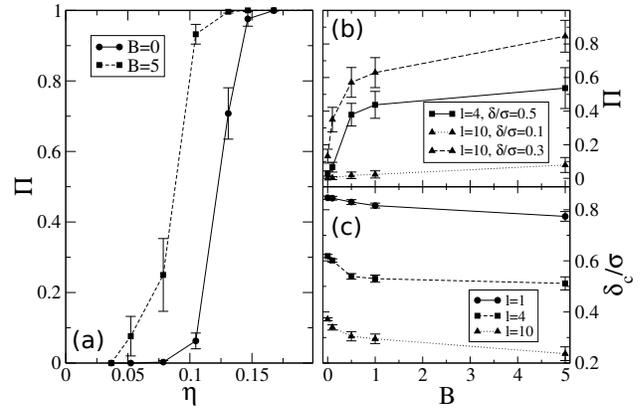}
%\vspace{1cm}
  \caption{\label{fig:PivB} (a) Percolation probability vs. volume fraction for $l=10$ and $m=1.5$ using $\delta=0.1\sigma$. (b) Percolation probability vs. field 
	   strength for different values of $l$ and $\delta$. (c) The parameter $\delta_c$ vs. field strength. In parts (b) and (c), the packing fraction $\eta=0.0524$.}
\end{figure}

A further interesting question is to which extent the external field influences the
percolation phase transition. Some representative results in this context
are shown in Fig.~\ref{fig:PivB}. Specifically, in Fig.~\ref{fig:PivB}(a) we have plotted
the percolation probability in a field of strength $B=5$ as a function of $\eta$ for rods of
length $l =10$ and dipole moment $m=1.5$. Comparing this curve
with the corresponding zero-field result [$B=0$, also included in Fig.~\ref{fig:PivB}(a)] we find that the percolation
threshold is shifted towards lower packing fractions. In other words, the magnetic field
{\em supports} the percolation transition in this system. In Fig.~\ref{fig:PivB}(b) we present data
for the percolation probability $\Pi$ as function of $B$ at a fixed packing fraction.
An increase of $B$ enhances the percolation probability, consistent with our finding from Fig.~\ref{fig:PivB}(a).
One also sees from Fig.~\ref{fig:PivB}(b) that the actual value of $\Pi$ in presence of a field strongly depends on the
rod length (consistent with the zero-field situation) as well as on the the parameter $\delta$ (determining our cluster criterion).
To complement the discussion, we present in Fig.~\ref{fig:PivB}(c) results for the "critical" distance $\delta_c$ as function of the field and different rod length, $l$.
For all lengths studied the value of
$\delta_c$ decreases upon an increase of $B$. This suggests that
the percolation threshold is indeed reduced by the presence of an external field,
{\em regardless} of the specific choice of $\delta$.

Finally, it is worth to briefly comment on the particle structures observed in an external field.
One general observation was that the structures
of Type~II (antiparallel local ordering) disappear, as expected from our results for the magnetization [see Fig.~11].
At the same time, structures of Type~III become prevalent,
along with a certain amount of Type~I structures. The clusters
that percolate under the effect of the external field do so in the
direction of the field, that is, the system essentially percolates along one direction. This is in contrast to the zero-field situation, where the percolating
clusters tend
to span the simulation box in all three spatial directions. As a consequence, percolation in field
leads to a strongly anisotropic network. This feature might become important e.g. for the design of
systems with conductance anisotropy.

\section{Conclusions}\label{concl}
Based on MC computer simulations we have explored structure formation phenomena in systems of model magnetic nanorods composed of $l$ fused dipolar spheres. The particle model has been inspired by recent experiments \cite{Birringer}, where MNRs have been produced via a self-assembly process. 
Compared to the experimental systems, we note that the longest rods considered in our simulations ($l=10$) have about half the length
of their experimental counterparts ($l\approx 24$). This is because simulating longer rods would have required larger system sizes and thus much longer simulation (equilibration) times. Nevertheless, we are already in the realistic range. The same holds true for the range of reduced dipole moments ($m\leq 2.4$) used in our simulations, which
approaches the experimental value of $m\approx 2.6$ (for rods composed of iron nanospheres at room temperature \cite{Birringer}).  Concerning the packing fractions, on the other hand, only the lowest value considered here ($\eta\approx 0.05$) is close to current experiments \cite{Birringer} ($\eta\approx 0.01-0.02$). However, it is just the advantage
of computer simulations that one can explore much larger parameter ranges.

Our simulation results generally reveal that the MNRs behave strongly different
not only from systems of individual magnetic spheres, but also from non-magnetic rods or rod-like particles with {\em single} dipoles.
Most of our results refer to systems in zero field. Compared to the case of magnetic spheres with central point dipoles ($l=1$),
we have found that the percolation threshold for $l>1$ is lowered towards significantly lower packing fractions. At the same
time, the percolation thresholds are also much lower than those of non-magnetic rods of comparable lengths.
These findings indicates that the MNRs could be promising candidates as building blocks of lightweight nanocomposites, i.e., connected materials
with novel mechanical and conductivity properties.
Moreover, for sufficiently large $l$ and densities beyond the percolation threshold, the MNRs display an isotropic-to-nematic transition, but no long-range ferromagnetic ordering, which is again
in contrast to magnetic spheres. These differences regarding the percolation and ordering behavior can be explained, on a microscopic level, from the fact that the present MNRs
display a larger variety of cluster structures. In particular, in addition to the usual head-to-tail ordering of dipolar spheres, we observe formation
of local antiparallel ordering, as well as of parallel side-by-side ordering with {\em shifted} positions. 

The presence of the latter type of clusters also represents one main difference 
to systems of rod-like particles with single (central, longitudinal) dipole moments. Indeed, this simpler model mainly exhibits
compact clusters with local antiparallel ordering \cite{GilJacksonMcGrother1}. As a consequence, these single-dipolar rods rather exhibit smectic (instead of nematic) phases \cite{Williamson,GilJacksonMcGrother1}. From a more general perspective, our simulations
results therefore suggest that the distribution of dipoles within the nanorods is of crucial importance for both, their percolation
behavior and for the nature of ordered phases. 

Finally, we have briefly discussed the impact of a static magnetic field. Our main conclusion in this context is
that already relatively weak fields (with a strength of about $2$~mT) can again significantly lower the
percolation threshold compared to the zero-field situation.

In the present paper we were concerned with the (interaction- or field-driven) structural behavior of the MNRs in thermal equilibrium. 
From an experimental point of view, it would be very interesting to explore the impact of these structural phenomena
on the single-particle dynamics (e.g., the translational and rotational diffusion) and the rheological properties of the suspension
(in particular its shear viscosity). Clearly, an investigation of these phenomena requires to supplement the MC simulations employed in the present work by computer simulations
targeting the time-dependent behavior, such as Brownian dynamics. 
Work in these directions is in progress.

\footnotesize{
\bibliography{rsc} %your .bib file
\bibliographystyle{rsc} %the RSC's .bst file
}

\end{document}